\documentclass[12pt]{article}
\usepackage[cm]{fullpage}
\usepackage{authblk}
\usepackage[superscript]{cite}

\usepackage[labelsep=period]{caption}

\usepackage{graphicx}
\usepackage{tikz}
\usetikzlibrary{decorations.pathmorphing}

\usepackage[version=3]{mhchem}
\usepackage{textcomp}

\title{Size Effect on the Short Range Order in Amorphous Materials}
\author[1]{Leonid Bloch}
\author[1]{Yaron Kauffmann}
\author[1,2]{Boaz Pokroy\thanks{bpokroy@tx.technion.ac.il}}
\affil[1]{Department of Materials Science and Engineering, Technion -- Israel Institute of Technology, Haifa, Israel.}
\affil[2]{Russell Berrie Nanotechnology Institute, Technion -- Israel Institute of Technology, Haifa, Israel.}
\date{\vspace{-2em}}

\begin{document}
\maketitle

\begin{abstract}
Drawing inspiration from nature, where some organisms can control the
short range order of amorphous minerals, we successfully manipulated
the short range order of amorphous alumina by surface and size effects.
By utilizing the Atomic Layer Deposition (ALD) method to grow
amorphous nanometrically thin films,
combined with state-of-the-art electron energy
loss spectroscopy (EELS) and X-ray photoelectron spectroscopy (XPS),
we showed experimentally that the short range order in such films is
strongly influenced by size.
This phenomenon is equivalent to the well-known size effect on
lattice parameters and on the relative stability of different
polymorphs in crystalline materials.
We also show that the short range order changes while still in the
amorphous phase, before the amorphous to crystalline transformation
takes place.
\end{abstract}

\section{}\label{main}
Many organisms use crystallization through an amorphous phase as the
route for biologically controlled mineralization\cite{biomin1,biomin2}.
This approach has several advantages, such as lowering the activation
energy for crystal formation\cite{kinetic-bio}, control over the final
shape of the crystal, controlled incorporation of impurities,
and control over the resulting polymorph\cite{stabil-CaCO3}.
The latter is achieved by manipulation of the short range order in the
amorphous phase, to resemble that of the succeeding crystalline
phase\cite{biomin3}.
This control is achieved through different additives, such as organic
molecules, water, magnesium, or phosphorus\cite{biomimetic-gen}.

This strategy used by organisms, if emulated synthetically, would
have a tremendous impact, as many aspects of science and technology rely
on amorphous materials\cite{amorphous-gen,amorphous-book1} and require
specific crystalline structures
in crystalline materials\cite{Emil-crystallography}.
The crystalline structure can be controlled mainly through temperature
and pressure manipulation (thermodynamics) and by
epitaxial growth\cite{Crystal_Engineering}.
However, to the best of our knowledge there is no study or technology
in which a polymorph has been stabilized via control of the short range order
of its amorphous precursor in a non-biological system.  

In our quest to emulate this biological strategy, we note
the well-known feature of nanometric scale structures,
namely that the influence of surface properties on the bulk is very
significant\cite{size,size-thermo}. It was shown, for example, that size
influences the lattice parameters of crystalline
nanoparticles\cite{surface-gold,GoldNano} and changes the
comparative stability of different
phases\cite{Navrotsky-alumina1,zirconia-micro}.
This occurs due to the increasing influence of
surface stress and surface energy as the particle size decreases.
This influence does not require the particle to be
crystalline, but should be present in any solid material\cite{SurfaceStress}.
 
In the present study we investigated whether the short range order of
an amorphous material can indeed be manipulated solely via size.
We chose atomic layer deposition (ALD) as our material deposition method,
since it is a technique that can provide extremely precise,
sub-nanometric, thickness control and can deposit conformal and
pinhole-free amorphous films of various materials\cite{ALD-overview}.
One of the most common processes for which the ALD is used
is the growth of thin, amorphous, \ce{Al2O3} films, mainly for the micro-
and nano-electronics industry\cite{ALD-overview}.
Therefore, our material of choice in this study was amorphous alumina.
An additional advantage of using \ce{Al2O3} is that it has many metastable
polymorphs at room temperature\cite{Levin-Brandon},
which may also suggest an ease of switching between different short
range order states.

\begin{figure}[!htb]
\centering
  \includegraphics[width=0.48\textwidth]{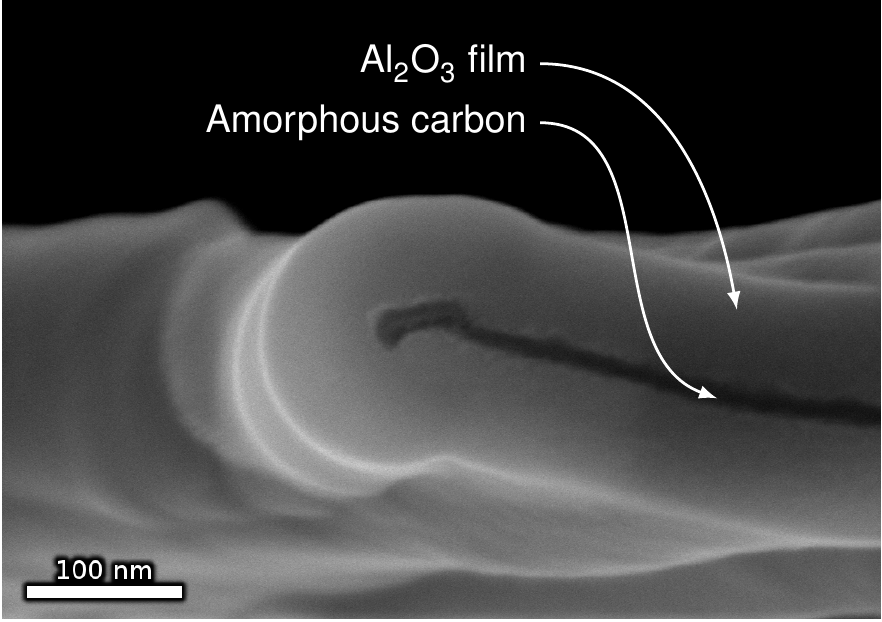}
  \caption{\textbf{Cross-section of coated TEM grid.} This
  scanning electron microscope (SEM) micrograph of \ce{Al2O3}
  film deposited on a holey carbon TEM grid shows the conformity of
  the film. The area seen in the image is an edge of one of
  the holes in the carbon film.}
  \label{fig:grid}
\end{figure}

Investigating the short range order of amorphous materials is more complex
than studying the crystalline structure of crystalline materials.
In this study we chose to use transmission electron microscopy (TEM)
and in particular the electron energy loss spectroscopy EELS method,
as well as X-ray photoelectron spectroscopy (XPS), to probe the short
range order. We deposited various nanometric thicknesses of amorphous
alumina directly onto `holey' amorphous carbon and amorphous \ce{Si3N4}
membrane TEM grids. After deposition the samples were transferred directly
into the TEM, and EELS was performed without any further sample manipulation
or preparation. An example of such a film deposited on a holey amorphous
carbon membrane can be seen in Fig. \ref{fig:grid}. It is apparent that
the film, as expected, is very conformal and void-free.

\begin{figure}[!htb]
\centering
  \includegraphics[width=0.48\textwidth]{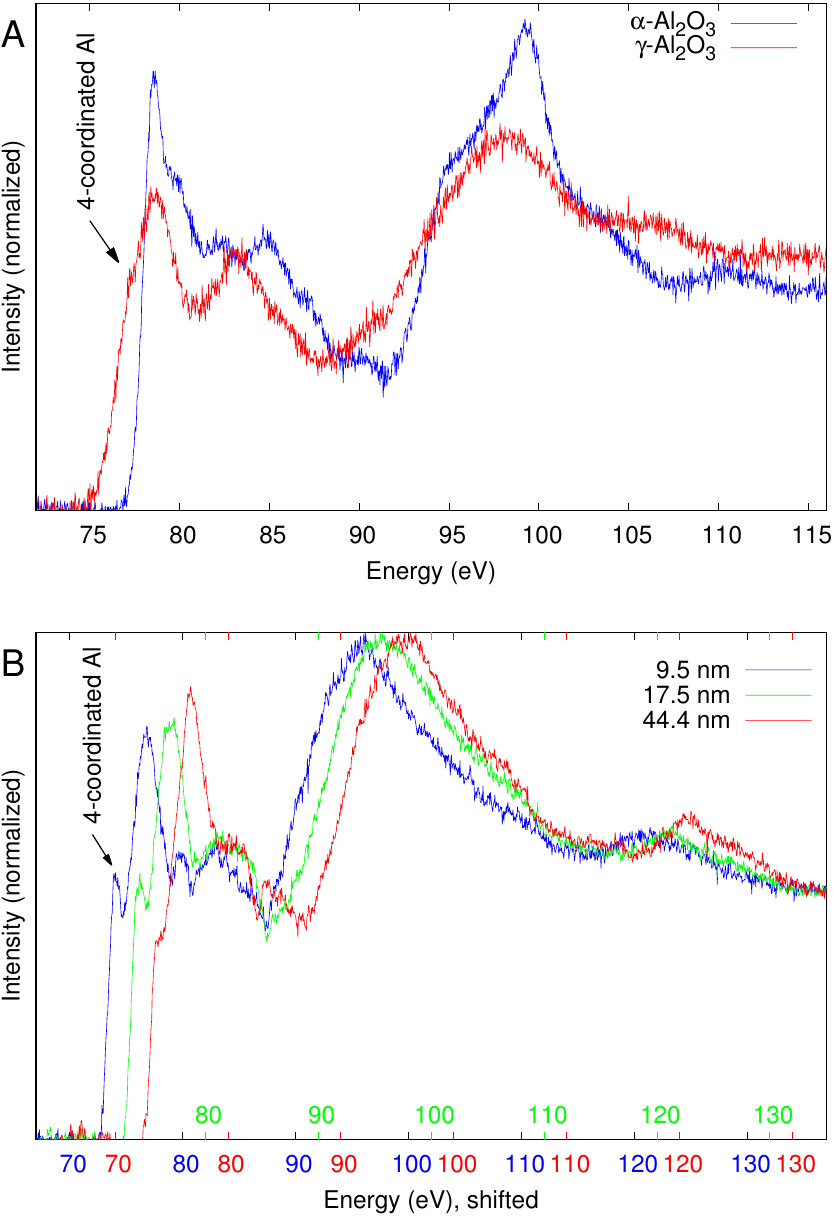}
  \caption{\textbf{Comparative EELS measurements, Al L$_{2,3}$-edge.}
  All spectra
  are background subtracted and 
  are normalized by integrated intensity. The arrow indicates the shoulder
  feature caused by the presence of tetrahedral Al sites. (\textbf{A})
  $\alpha$ and $\gamma$-\ce{Al2O3} nano-powder references. (\textbf{B})
  Measurements on films of three different thicknesses.
  The spectra are shifted slightly to emphasize the differences
  (hence ``shifted'' on the energy axis label).}
  \label{fig:EELS-comp}
\end{figure}

The deposited films were amorphous, which could be verified by electron
diffractions (fig. \ref{supp:diff-before}).
EELS reference measurements were performed on sub-$\mu m$ powders
of crystalline $\alpha$-\ce{Al2O3} and $\gamma$-\ce{Al2O3}, at the
aluminum L$_{2,3}$-edge (Fig. \ref{fig:EELS-comp}A).
The resulting spectra conformed well with
literature reports\cite{Levin-EELS,thermal-annealing-Al2O3}.
It should be noticed that a pre-edge shoulder feature at the aluminum
L$_{2,3}$-edge can be observed only in the case of $\gamma$-\ce{Al2O3}.
This is because of the presence of 4-coordinated (tetrahedral) Al
sites\cite{Levin-EELS,EELS-alumina-calc}, which are
present only in the $\gamma$-\ce{Al2O3} structure, and not in that of
$\alpha$-\ce{Al2O3}, in which all the Al sites are rather
6-coordinated (octahedral).

Similar measurements were performed on the ALD-deposited amorphous films
of different thicknesses (Fig. \ref{fig:EELS-comp}B). A clear
trend could be observed in these measurements, namely that the
prominent 4-coordinated Al feature (shoulder appearance at the Al
L$_{2,3}$-edge) is more pronounced in the case of thinner films,
whereas for the thicker ones (above approximately 20 nm) the shoulder
that appears in the spectrum is much weaker.
This means that thicker amorphous films demonstrate, on average,
a short range order in which fewer tetrahedral Al sites are present,
while thinner amorphous films demonstrate a structure richer in these sites.
This finding is striking, as the only difference between these films is
their thickness (deposition parameters were kept constant for all samples).

\begin{figure*}[!htb]
\centering
  \includegraphics[width=1.0\textwidth]{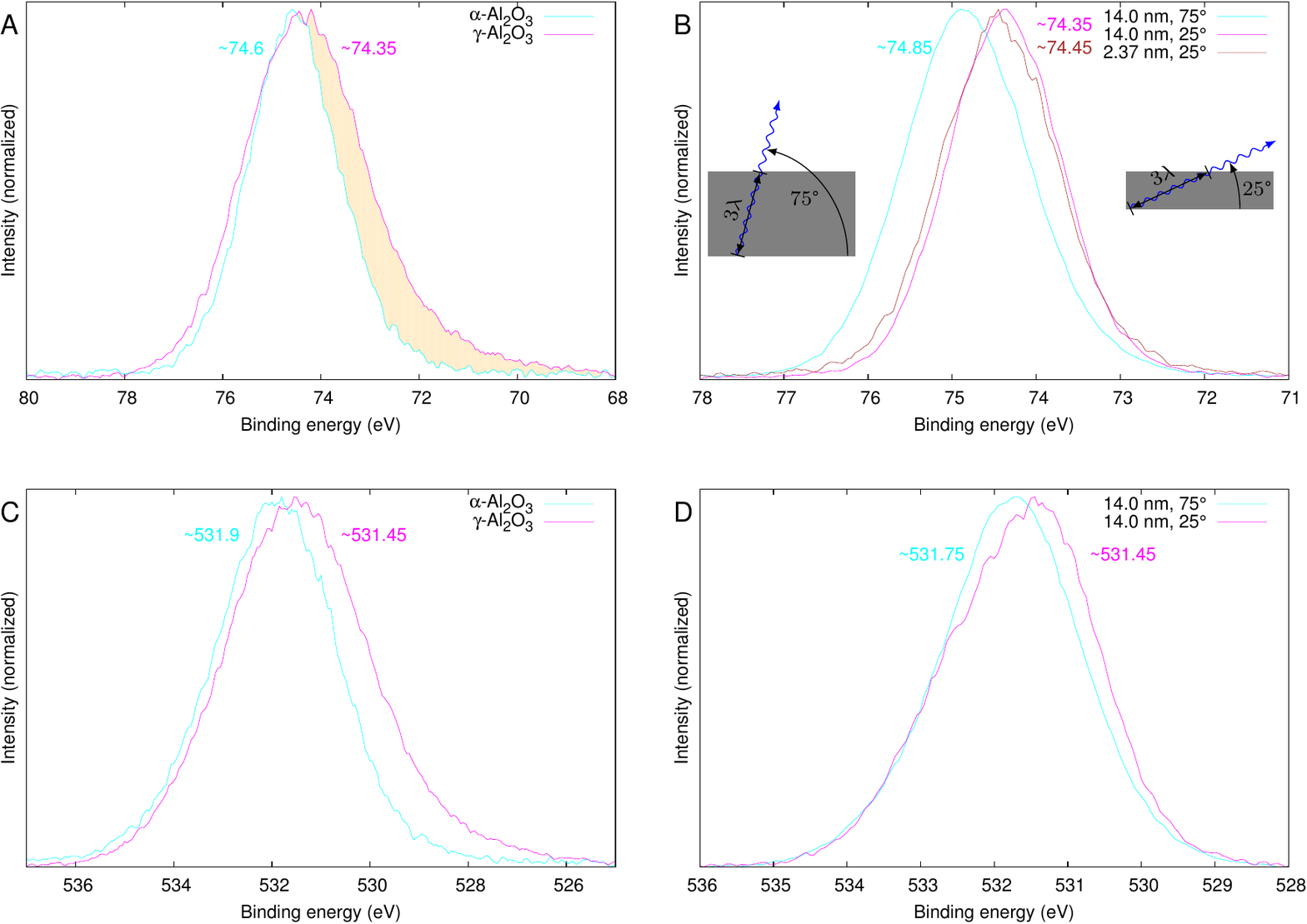}
  \caption{\textbf{XPS measurements, Al2p and O1s peaks.}
  (\textbf{A}) Al2p spectra of crystalline \ce{Al2O3} references.
  A shift to lower energies can be seen for $\gamma$-\ce{Al2O3},
  due to the presence of 4-coordinated Al.
  (\textbf{B}) Al2p spectra from two different photoelectron take-off
  angles of a 14 nm thick amorphous \ce{Al2O3} film, and
  a 2.37 nm thick amorphous film. A noticeable shift to lower
  energies can be seen for the lower take-off angle and the thin sample.
  (\textbf{C}) O1s spectra of crystalline \ce{Al2O3} references.
  A shift to lower energies can be seen for $\gamma$-\ce{Al2O3}.
  (\textbf{D}) O1s spectra from two different photoelectron take-off
  angles of a 14 nm thick amorphous \ce{Al2O3} film. A shift to lower
  energies is visible for the lower take-off angle.}
  \label{fig:XPS}
\end{figure*}

To further verify the EELS measurements, we used the XPS technique.
A reference measurement was performed on the same powders of crystalline
$\alpha$ and $\gamma$ alumina as those on which the EELS calibration
measurements were performed. Fig. \ref{fig:XPS}A shows the Al2p peaks
acquired by this measurement. Peak broadening can be seen, as well as
a slight shift towards lower energies in the case of $\gamma$-\ce{Al2O3},
relatively to the peak of $\alpha$-\ce{Al2O3},
in good agreement with the EELS results 
(see Fig. \ref{fig:EELS-comp}A). 
This is attributable to the presence in $\gamma$ alumina of
tetrahedral Al sites, which have lower binding energy than the octahedral Al
sites\cite{XPS-Gonzalez-Elipe,XPS-Ebina}.

To investigate the size effect on amorphous thin films we had to
take into consideration that in contrast to the EELS method, in which the
signal originates from the entire sample thickness, in XPS the signal
originates only from the top several nanometers below the surface.
We therefore performed angle-resolved XPS (ARXPS), which enabled us to
analyze the sample with depth resolution, by sampling photoelectrons from
different takeoff angles. The sampling depth from which 95\% of the signal
intensity in ARXPS comes can be roughly estimated as $3\lambda\sin\theta$,
where $\lambda$ is the attenuation length for photoelectrons in the sample,
and $\theta$ is the takeoff angle of the photoelectrons\cite{ARXPS}.
Thus, the sampling depth increases when probing at higher takeoff
angles (see inserts in Fig. \ref{fig:XPS}B).

For this procedure we used thin amorphous alumina films with the same
deposition parameters as those used for the EELS samples, but these films
were deposited on silicon wafers, rather than on the TEM grids. We used
two film thicknesses, 2.37 nm and 14 nm. Two takeoff angles were
chosen, 25{\textdegree} and 75{\textdegree}, which, for alumina,
correspond to sampling depths of roughly 3.6 nm and
8.1 nm\cite{NIST-attenuation}, respectively.

We first investigated the Al2p peak of the 14 nm film at both takeoff
angles. In Fig. \ref{fig:XPS}B, a shift to lower binding energies from
the shallower takeoff angle (25{\textdegree}, $\sim$3.6 nm) is seen,
suggesting that the less energetic, 4-coordinated Al sites are more
abundant closer to the film surface. 
We then investigated the Al2p peak of the 2.37 nm film, at the
25{\textdegree} takeoff angle only, as at this angle the signal already
originates from the entire film thickness. In both of the 25{\textdegree}
measurements, on the thicker film as well as on the thinner one, the
majority of the signal is derived from roughly the same sample thickness,
and it appears that there is practically no shift in the
binding energies between them.
These observations, like the EELS measurements, indicate that the
fraction of tetrahedral Al sites is greater near the surface of
amorphous films and hence also in thinner amorphous films.

We performed similar experiments, on the same samples,
to investigate the oxygen binding energy, by probing the O1s peak.
In Fig. \ref{fig:XPS}, C and D, a similar trend can be observed for this
peak. The signal from shallower depth is shifted slightly to lower
binding energies, similarly to the shift of the O1s signal from
$\gamma$-\ce{Al2O3}, relatively to $\alpha$-\ce{Al2O3}.
This provides further evidence that the short range order near the
surface is more ``$\gamma$-like'' than ``$\alpha$-like''.
The 2.37 nm thin film sample could not be examined for binding energy
of oxygen sites, as the oxygen from the silica substrate would
interfere. On the other hand, we made sure that no such interference was
present in the 14 nm sample, by verifying that no Si peaks were visible
in the spectra from either of the takeoff angles (fig. \ref{supp:XPS-noSi}).

\begin{figure*}[!htb]
\centering
  \includegraphics[width=0.59\textwidth]{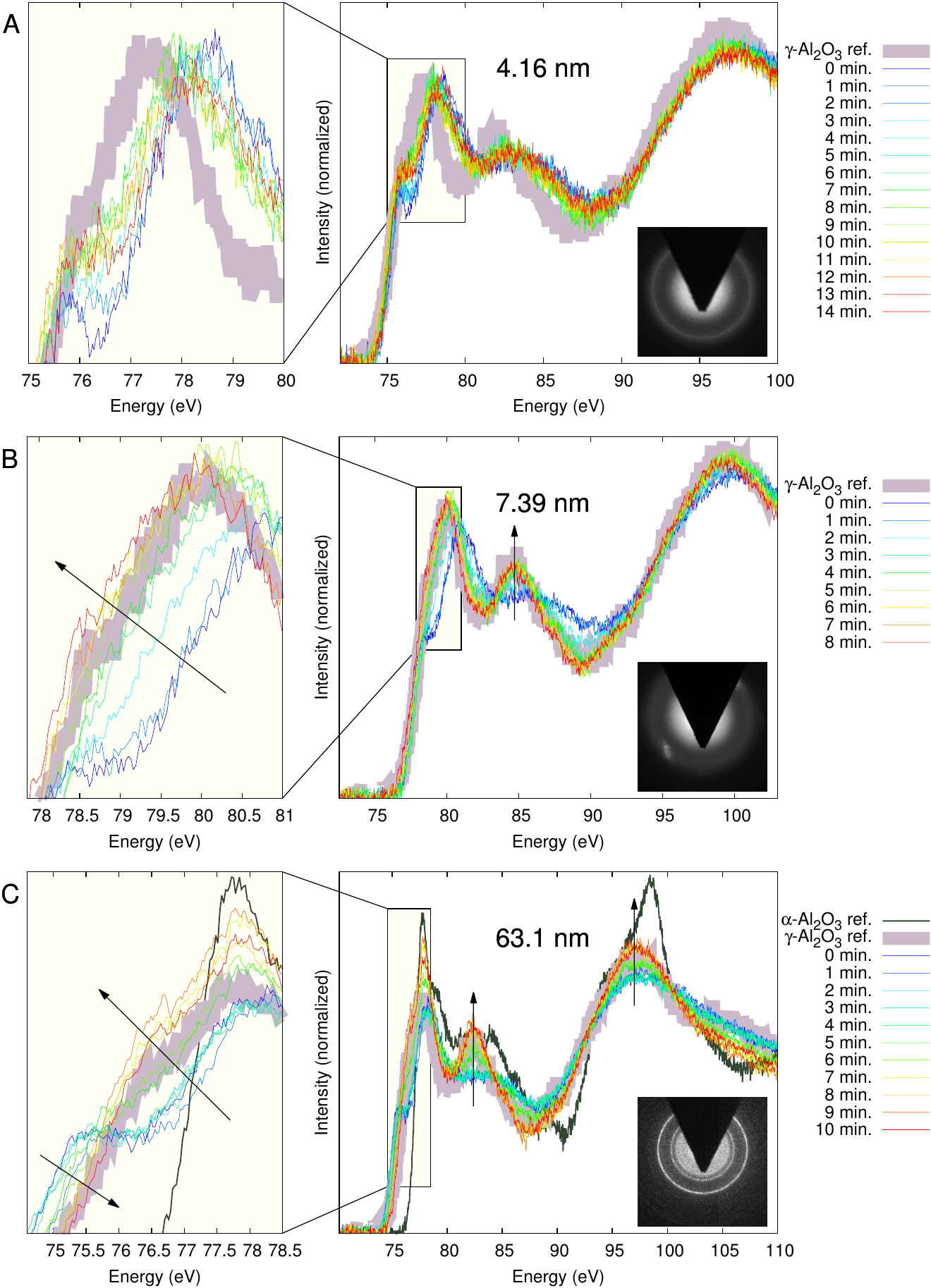}
  \caption{\textbf{Evolution of EELS spectra under electron beam,
                   Al L$_{2,3}$-edge.}
  The times specify the duration of exposure of samples to the
  electron beam, not including the measurement time.
  All spectra are presented as obtained, without any manipulation,
  apart from background subtraction.
  Only the $\gamma$-\ce{Al2O3} spectrum is shifted between figures in
  order to be aligned with the time resolved spectra. The diffraction
  shown in each figure was taken after all the measurements were finished,
  from the area on which they were performed.
  (\textbf{A}) 4.16 nm thick film: only a slight change is observed,
  the shoulder feature emphasized in Fig. \ref{fig:EELS-comp} rises
  slightly, and then fluctuates around the same point.
  No crystallization occurs.
  (\textbf{B}) 7.39 nm thick film: after several minutes the spectrum
  starts to repeat almost perfectly the shape of the reference
  $\gamma$-\ce{Al2O3} spectrum. The shoulder feature caused by
  tetrahedral Al sites is smoothed. Some crystallization occurs.
  (\textbf{C}) 63.1 nm thick film: on this film, some
  $\alpha$-\ce{Al2O3}--like features begin to appear, the shoulder
  feature is smoothed from both directions, and obvious crystallization
  is noticeable after the measurements.}
  \label{fig:EELS-time}
\end{figure*}

Once it became clear that the short range order does indeed depend on the
thickness of the amorphous film, and having noticed that some of these
amorphous films crystallized after being exposed for some time
(about 15 minutes) to the electron beam in the TEM, we wanted to further
investigate how the short range order evolves, while still in the amorphous
state, during exposure to the electron beam and just before the amorphous
to crystalline transformation. To this end, time-dependent EELS
measurements were performed on amorphous alumina films of varying
thicknesses (Fig. \ref{fig:EELS-time}). A constant area of each sample
was exposed to a focused electron beam (with the same parameters between
different samples) and EELS measurements on Al L$_{2,3}$-edge were
performed with one minute intervals between them.
The calibration was the same for all samples. 

For the thinnest sample (4.16 nm), Fig. \ref{fig:EELS-time}A clearly
shows that fast rising and some smoothing of the initially extremely
prominent shoulder feature occurred, but afterwards,
this feature continued oscillating around the same point,
and overall, there were hardly any changes in other features of the spectrum.
Even after 20 minutes of exposure and measurements, the short range order did
not stabilize on a specific state, but kept fluctuating back and forth.
After the time-resolved measurement we switched to diffraction mode, and
found that the sample remained in its amorphous state, and did not crystallize
(see insert in Fig. \ref{fig:EELS-time}A).

In another film, which was just $\sim$3 nm thicker
(7.39 nm; Fig. \ref{fig:EELS-time}B), the behavior was different:
smoothing of the shoulder feature was continuous, and simultaneously,
another peak started appearing and increasing in intensity at approximately
84.7 eV. Interestingly, it can be seen that the spectrum continuously moves
towards resembling that of the $\gamma$-\ce{Al2O3}, until at some point it
overlaps it perfectly. Here we want to stress the remarkable finding that
the short range order evolved continuously towards that of the
$\gamma$-\ce{Al2O3}, even before crystallization took place.
At the end of this measurement, we again switched to diffraction mode,
which took several minutes while the beam was still focused on the same area,
and noticed a very strange pattern (see insert in Fig. \ref{fig:EELS-time}B).
No diffraction rings could be seen, however at one azimuthal angle a
clustered scatter of points appeared. This is very surprising, as it implies
that the sample crystallized under the beam at a highly preferred orientation.
Even more surprising is the fact that this pattern has a wide
range of d-spacings. At this point we cannot fully explain this phenomenon,
but we can state that at some point we probably did, locally, crystallize the
film.      

Finally, we similarly investigated a considerably thicker sample (63.1 nm).
In this case we again noticed a time-resolved change in the short range
order: the shoulder feature decreased considerably,
and the spectrum exhibited some $\alpha$-\ce{Al2O3}--like features
with time (Fig. \ref{fig:EELS-time}C). It should be pointed out, however,
that in this case we did not get a perfect overlap between the final spectrum
and any of the reference spectra. The final spectrum was close to that
of $\gamma$-\ce{Al2O3}, despite the appearance of few
$\alpha$-\ce{Al2O3}--like features. In this case, switching to diffraction
mode clearly revealed polycrystalline diffraction rings attributable
to $\gamma$-\ce{Al2O3} (fig. \ref{supp:450-diff}), but there was no
sign of crystalline $\alpha$-\ce{Al2O3}.

These results clearly show that the short range order in an amorphous
material does change as a function of thickness (size).
This is of great importance, among other reasons because the short range
order in an amorphous material influences the
crystalline structure succeeding after it's crystallization.
The EELS analysis coincides with the XPS data, which show that the surface
of amorphous alumina possesses a different short range order than the
average in it's bulk, and the thinner, or smaller,
the amorphous solid is, the more it's short range order resembles that
near the surface.
This phenomenon is due to the stronger effect that the surface exerts on
the bulk as the size decreases. The surface structure we observed in our
study agrees with what was predicted by molecular dynamics
simulations\cite{simulations}.

This study demonstrates that the short range order in amorphous materials
can indeed be altered via size effects, just as the lattice parameters of
crystalline materials can be altered by them. We believe that these findings
will open up new areas of research, as well as drive towards the ability to
control the short range order of amorphous materials, where this is
a cardinal issue that could not be previously addressed, due to the lack of
possibility of such control.

\section{Acknowledgments}
The research that yielded these results received funding from the
European Research Council under the European Union's Seventh Framework
Program (FP/2007--2013)/ERC Grant Agreement {n\textsuperscript{o}} [336077].
We thank Dr. L. Burstein from the Wolfson Applied
Materials Research Center, Tel Aviv University, for the XPS measurements,
and Dr. I. Levin, Prof. W. Kaplan, Dr. O. Kreinin, and Dr. K. Jorissen for
helpful discussions.

\bibliographystyle{ieeetr}
\bibliography{main}

\begin{thebibliography}{10}

\bibitem{biomin1}
S.~Weiner and L.~Addadi, ``Crystallization pathways in biomineralization,''
  {\em Annual Review of Materials Research}, vol.~41, no.~1, pp.~21--40, 2011.

\bibitem{biomin2}
S.~Weiner, I.~Sagi, and L.~Addadi, ``Choosing the crystallization path less
  traveled,'' {\em Science}, vol.~309, no.~5737, pp.~1027--1028, 2005.

\bibitem{kinetic-bio}
P.~Fratzl, F.~D. Fischer, J.~Svoboda, and J.~Aizenberg, ``A kinetic model of
  the transformation of a micropatterned amorphous precursor into a porous
  single crystal,'' {\em Acta Biomaterialia}, vol.~6, no.~3, pp.~1001--1005,
  2010.

\bibitem{stabil-CaCO3}
J.~Aizenberg, L.~Addadi, S.~Weiner, and G.~Lambert, ``Stabilization of
  amorphous calcium carbonate by specialized macromolecules in biological and
  synthetic precipitates,'' {\em Advanced Materials}, vol.~8, no.~3,
  pp.~222--226, 1996.

\bibitem{biomin3}
L.~Addadi, S.~Raz, and S.~Weiner, ``Taking advantage of disorder: Amorphous
  calcium carbonate and its roles in biomineralization,'' {\em Advanced
  Materials}, vol.~15, no.~12, pp.~959--970, 2003.

\bibitem{biomimetic-gen}
L.~B. Gower, ``Biomimetic model systems for investigating the amorphous
  precursor pathway and its role in biomineralization,'' {\em Chemical
  reviews}, vol.~108, no.~11, pp.~4551--4627, 2008.

\bibitem{amorphous-gen}
D.~A. Drabold, ``Topics in the theory of amorphous materials,'' {\em The
  European Physical Journal B}, vol.~68, no.~1, pp.~1--21, 2009.

\bibitem{amorphous-book1}
S.~R. Elliott, {\em Physics of Amorphous Materials}.
\newblock Essex, UK: Longman, 2~ed., 6 1990.

\bibitem{Emil-crystallography}
E.~Zolotoyabko, {\em Basic Concepts of Crystallography}.
\newblock Weinheim, Germany: Wiley-VCH, 5 2011.

\bibitem{Crystal_Engineering}
D.~Braga, F.~Grepioni, and A.~G. Orpen, eds., {\em Crystal Engineering: From
  Molecules and Crystals to Materials}.
\newblock Springer, 9 1999.

\bibitem{size}
R.~Berry, ``Thermodynamics -- size is everything,'' {\em Nature}, vol.~393,
  pp.~212--213, MAY 21 1998.

\bibitem{size-thermo}
A.~Navrotsky, ``Nanoscale effects on thermodynamics and phase equilibria in
  oxide systems,'' {\em ChemPhysChem}, vol.~12, no.~12, pp.~2207--2215, 2011.

\bibitem{surface-gold}
C.~Mays, J.~Vermaak, and D.~Kuhlmann-Wilsdorf, ``On surface stress and surface
  tension: Ii. determination of the surface stress of gold,'' {\em Surface
  Science}, vol.~12, no.~2, pp.~134--140, 1968.

\bibitem{GoldNano}
C.~Solliard and M.~Flueli, ``Surface stress and size effect on the lattice
  parameter in small particles of gold and platinum,'' {\em Surface Science},
  vol.~156, Part 1, no.~0, pp.~487--494, 1985.

\bibitem{Navrotsky-alumina1}
J.~M. McHale, A.~Auroux, A.~J. Perrotta, and A.~Navrotsky, ``Surface energies
  and thermodynamic phase stability in nanocrystalline aluminas,'' {\em
  Science}, vol.~277, no.~5327, pp.~788--791, 1997.

\bibitem{zirconia-micro}
R.~C. Garvie, ``Stabilization of the tetragonal structure in zirconia
  microcrystals,'' {\em The Journal of Physical Chemistry}, vol.~82, no.~2,
  pp.~218--224, 1978.

\bibitem{SurfaceStress}
F.~Fischer, T.~Waitz, D.~Vollath, and N.~Simha, ``On the role of surface energy
  and surface stress in phase-transforming nanoparticles,'' {\em Progress in
  Materials Science}, vol.~53, no.~3, pp.~481--527, 2008.

\bibitem{ALD-overview}
S.~M. George, ``Atomic layer deposition: An overview,'' {\em Chemical Reviews},
  vol.~110, no.~1, pp.~111--131, 2010.
\newblock PMID: 19947596.

\bibitem{Levin-Brandon}
I.~Levin and D.~Brandon, ``Metastable alumina polymorphs: Crystal structures
  and transition sequences,'' {\em Journal of the American Ceramic Society},
  vol.~81, no.~8, pp.~1995--2012, 1998.

\bibitem{Levin-EELS}
I.~Levin, A.~Berner, C.~Scheu, H.~Muellejans, and D.~Brandon, ``Electron
  energy-loss near-edge structure of alumina polymorphs,'' in {\em Modern
  Developments and Applications in Microbeam Analysis} (G.~Love, W.~Nicholson,
  and A.~Armigliato, eds.), vol.~15 of {\em Mikrochimica Acta Supplement},
  pp.~93--96, Springer Vienna, 1998.

\bibitem{thermal-annealing-Al2O3}
V.~Edlmayr, T.~P. Harzer, R.~Hoffmann, D.~Kiener, C.~Scheu, and C.~Mitterer,
  ``Effects of thermal annealing on the microstructure of sputtered al2o3
  coatings,'' {\em Journal of Vacuum Science \& Technology A: Vacuum, Surfaces,
  and Films}, vol.~29, no.~4, pp.~041506--041506--8, 2011.

\bibitem{EELS-alumina-calc}
R.~Brydson, ``Multiple scattering theory applied to elnes of interfaces,'' {\em
  Journal of Physics D: Applied Physics}, vol.~29, no.~7, p.~1699, 1996.

\bibitem{XPS-Gonzalez-Elipe}
A.~R. Gonz\'{a}lez-Elipe, J.~P. Espin\'{o}s, G.~Munuera, J.~Sanz, and J.~M.
  Serratosa, ``Bonding-state characterization of constituent elements in
  phyllosilicate minerals by xps and nmr,'' {\em The Journal of Physical
  Chemistry}, vol.~92, no.~12, pp.~3471--3476, 1988.

\bibitem{XPS-Ebina}
T.~Ebina, T.~Iwasaki, A.~Chatterjee, M.~Katagiri, and G.~D. Stucky,
  ``Comparative study of xps and dft with reference to the distributions of al
  in tetrahedral and octahedral sheets of phyllosilicates,'' {\em The Journal
  of Physical Chemistry B}, vol.~101, no.~7, pp.~1125--1129, 1997.

\bibitem{ARXPS}
C.~S. Fadley, ``Angle-resolved x-ray photoelectron spectroscopy,'' {\em
  Progress in Surface Science}, vol.~16, no.~3, pp.~275--388, 1984.

\bibitem{NIST-attenuation}
C.~J. Powell and A.~Jablonski, {\em NIST Electron Effective-Attenuation-Length
  Database}.
\newblock Gaithersburg, MD: National Institute of Standards and Technology,
  2011.
\newblock Version 1.3.

\bibitem{simulations}
S.~P. Adiga, P.~Zapol, and L.~A. Curtiss, ``Atomistic simulations of amorphous
  alumina surfaces,'' {\em Phys. Rev. B}, vol.~74, p.~064204, Aug 2006.

\end{thebibliography}

\newpage

\section{Supplementary Materials}
\setcounter{figure}{0}
\makeatletter 
\renewcommand{\thefigure}{S\@arabic\c@figure}
\makeatother

\subsection{S1. Materials and Methods}
\paragraph{Atomic Layer Deposition (ALD):}
The ALD device used for these experiments was R-200
(Picosun, Finland). The growth process of \ce{Al2O3} films utilized
trimethylaluminum (TMA) and \ce{H2O} as precursors for deposition on
Si wafers and TEM grids. The thickness of the films was determined
by ellipsometry, directly on the Si wafer sample or, for other substrates,
on a piece of Si wafer that was placed close to the sample during
the growth process for this purpose.

\paragraph{Scanning Electron Microscopy (SEM):}
The image in Fig. \ref{fig:grid} was obtained with the Zeiss Ultra Plus
High Resolution FEG-SEM (Zeiss, Germany), operated at 4.0 kV,
using an in-lens secondary electrons (SE) detector.

\paragraph{Electron Energy Loss Spectroscopy (EELS):}
An FEI Titan 80--300 S/TEM (FEI, Eindhoven, The Netherlands) was operated
at 300~KeV and was equipped with an image Cs corrector. A post-column
energy filter (Tridiem 866 ERS, Gatan, USA) was used to obtain the
EELS measurements. The measured energy spread of the beam was
0.6 eV, and a dispersion of 0.03 eV/channel was used.

In order to avoid the influence of sample exchange on the spectra,
several coated grids were cut and placed on a single sample holder
for the comparative measurements.

\paragraph{X-ray Photoelectron Spectroscopy (XPS):}
XPS measurements were performed in UHV ($2.5~\times~10^{-10}$ Torr
base pressure) using the 5600 Multi-Technique System (PHI, USA).
Samples were irradiated with an Al K$_\alpha$ monochromated source
(1486.6 eV) and the outcome electrons were analyzed by a spherical
capacitor analyzer using a slit aperture of 0.8 mm.
Sample charging during measurements was compensated by means of a
neutralizer, with additional mathematical shift used when necessary
(C1s at 285 eV was taken as an energy reference for all the measured
peaks). The high resolution measurements presented in
figure \ref{fig:XPS} were taken in a low energy range window with
pass energy (PE) of 11.75 eV and 0.05 eV/step.
\newpage

\subsection{S2. Figures}
\begin{figure*}[!htb]
\centering
  \includegraphics[width=1.0\textwidth]{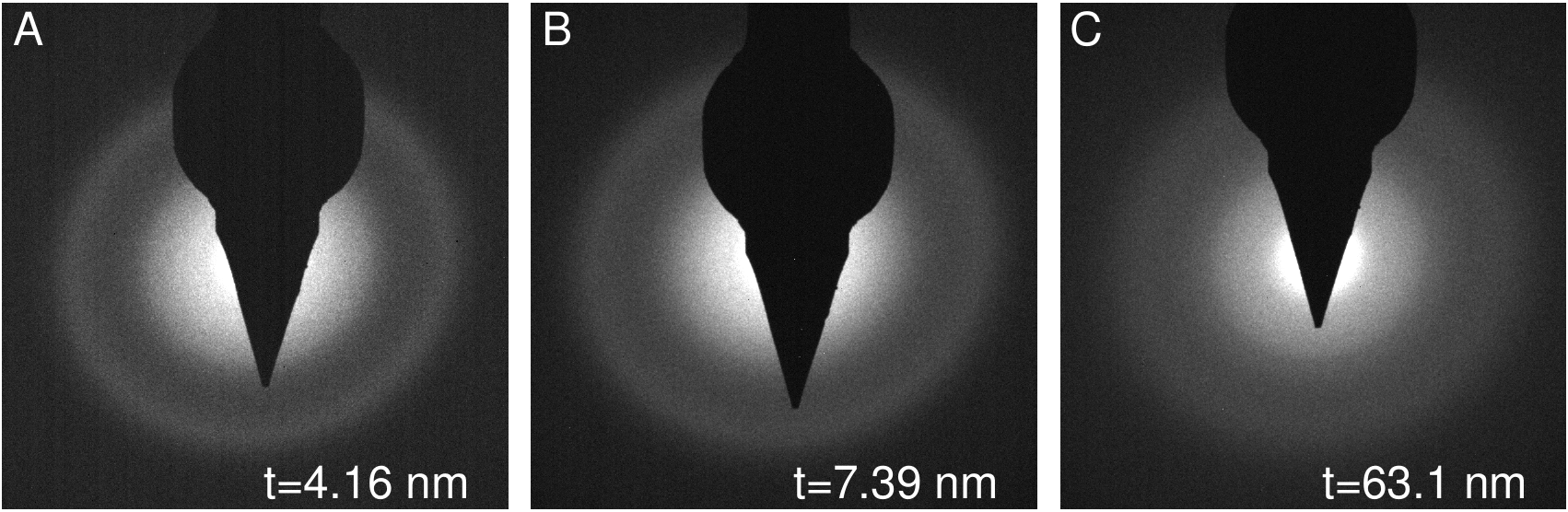}
  \caption{\textbf{Initial electron diffractions from the three samples
  used in the time-resolved EELS measurements that are presented in
  Fig. \ref{fig:EELS-time}.}
  The initial diffractions could not be taken from the same spots
  on which the time-resolved measurements were performed,
  as this would affect the results. Instead they were taken later,
  after all measurements were completed, from a nearby spot on the
  same samples.
  (\textbf{A})
  4.16 nm thick sample.
  (\textbf{B})
  7.39 nm thick sample.
  (\textbf{C})
  63.1 nm thick sample. It can be clearly seen that the as-deposited
  \ce{Al2O3} on all samples is amorphous.}
  \label{supp:diff-before}
\end{figure*}

\begin{figure*}[!htb]
\centering
  \includegraphics[width=0.66\textwidth]{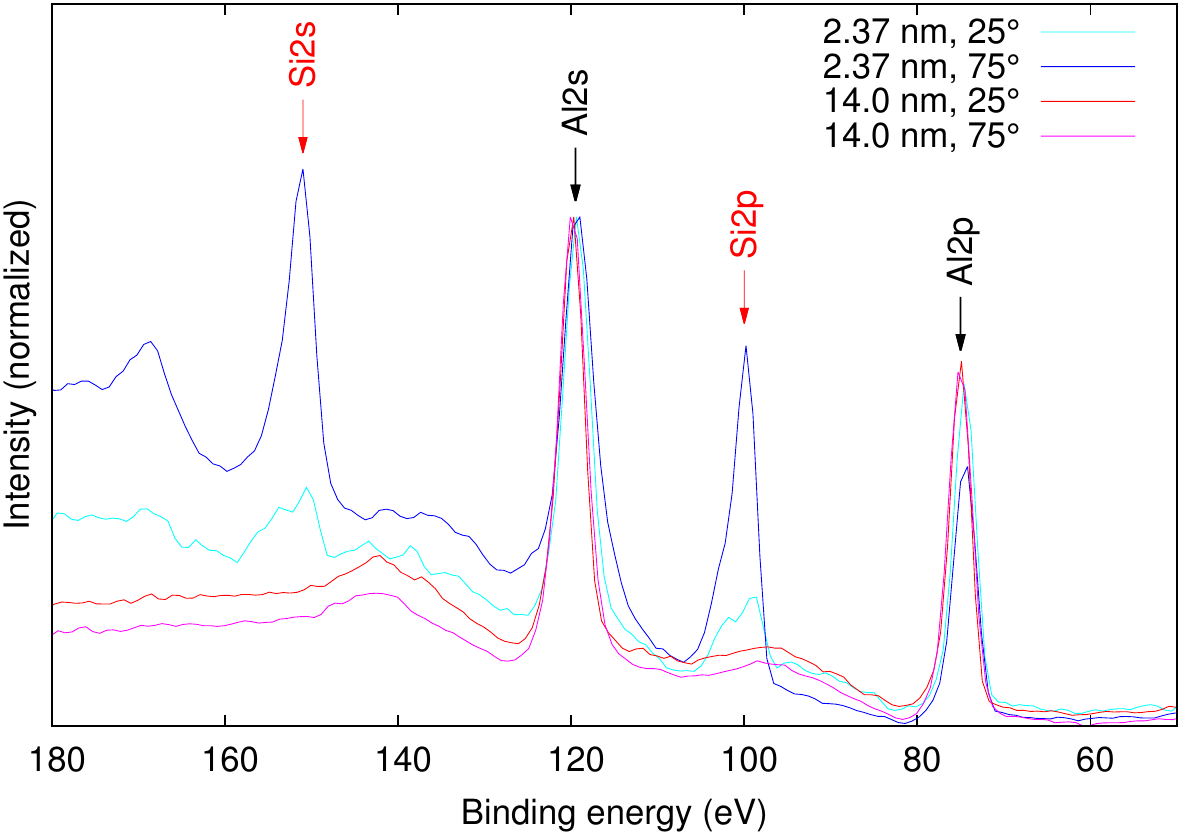}
  \caption{\textbf{XPS survey scan in the 180--50 eV range on the samples
  used in Fig. \ref{fig:XPS}.}
  The Si peaks can be clearly seen in the spectra of the thin (2.37 nm)
  sample from both takeoff angles, and are absent from the thick (14 nm)
  sample spectra. From this, we deduce that no signal is coming from the
  \ce{SiO2}/\ce{Al2O3} interface of the thick sample, whereas it does come
  from the thin sample, and therefore the oxygen signal from the thin
  alumina film would be convoluted with the oxygen signal from the
  underlying silica.}
  \label{supp:XPS-noSi}
\end{figure*}

\begin{figure*}[!htb]
\centering
  \includegraphics[width=0.48\textwidth]{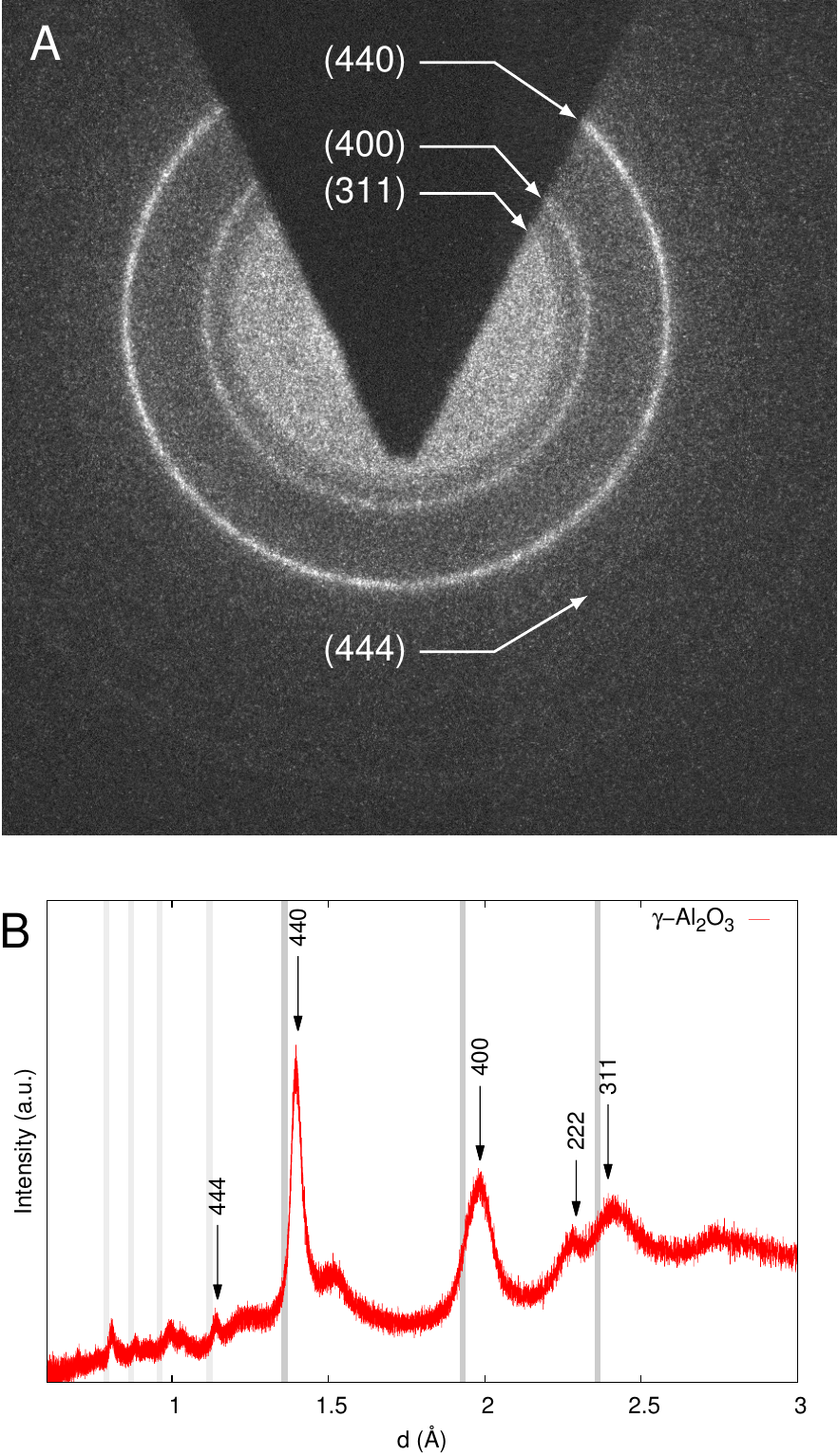}
  \caption{\textbf{Electron diffraction obtained from the 63.1 nm
  thick sample after the time-resolved measurements on it were completed.}
  Several minutes after completing the EELS measurements,
  this diffraction was taken from the spot on which they were performed.
  (\textbf{A})
  Miller indices corresponding to the crystalline planes of
  $\gamma$-\ce{Al2O3} are presented.
  (\textbf{B})
  X-ray diffraction (XRD) spectrum of $\gamma$-\ce{Al2O3} powder,
  the same powder as the one used for EELS and XPS reference spectra
  (XRD was performed at the ID31 beamline, European Synchrotron Radiation
  Facility (ESRF), Grenoble, France).
  Gray vertical stripes represent the rings from the
  electron diffraction above. Good resemblance can be seen, both to the
  peak locations and to their intensities. A slight shift to smaller
  d-spaces is evident in the electron diffraction. This might be due
  to a size effect on the lattice parameters (there are probably size
  differences between the crystallites in the powder and on the sample),
  and/or due to inaccuracy in the microscope calibration that was used to
  calculate the d-spaces from the electron diffraction. The darker gray
  stripes correspond to brighter rings, and the lighter stripes -- to
  the barely visible rings that appear on the diffraction image.}
  \label{supp:450-diff}
\end{figure*}

\end{document}